\documentclass[journal]{IEEEtran}
\usepackage{graphicx}

\begin{document}

\title{ACH: Away Cluster Heads Scheme for Energy Efficient Clustering Protocols in WSNs}

\author{N. Javaid$^{\ddag}$, M. Waseem$^{\ddag}$, Z. A. Khan$^{\$}$, U. Qasim$^{\pounds}$, K. Latif$^{\ddag}$, A. Javaid$^{\natural}$\\\vspace{0.4cm}
        COMSATS Institute of IT, $^{\ddag}$Islamabad, $^{\natural}$Wah Cantt, Pakistan. \\
        $^{\$}$Faculty of Engineering, Dalhousie University, Halifax, Canada.\\
        $^{\pounds}$University of Alberta, Alberta, Canada.
     }

\maketitle

\begin{abstract}

This paper deals with the routing protocols for  distributed wireless sensor networks. The conventional protocols for WSNs like Low Energy adaptive Clustering Hierarchy (LEACH), Stable Election Protocol (SEP), Threshold Sensitive Energy Efficient Network (TEEN), Distributed Energy Efficient Clustering Protocol (DEEC) may not be optimal. We propose a scheme called Away Cluster Head (ACH) which effectively increases the efficiency of conventional clustering based protocols in terms of stability period and number of packets sent to base station (BS). We have implemented ACH scheme on LEACH, SEP, TEEN and DEEC. Simulation results show that LEACH-ACH, SEP-ACH, TEEN-ACH and DEEC-ACH performs better than LEACH, SEP, TEEN and DEEC respectively in terms of stability period and number of packets sent to BS. The stability period of the existing protocols prolongs by implementing ACH on them.
\end{abstract}

\begin{keywords}
Wireless sensor networks, Distributed networks, Clustering Protocol.
\end{keywords}

\section{Background}

In Direct Transmission [1], each node in the sensor network communicates directly to BS. In the aforementioned protocol, farthest nodes die faster than the nearest nodes. In Minimum transmission energy [2] routing protocol each node transmits to its nearest node so the nearest nodes die at a faster rate because they receive data from the farther nodes. In the current body of research going in the field of WSNs clustering based protocols have attain significant attraction. In clustering based routing protocols the sensor nodes form clusters. In these clusters, one node is selected as CH. The nodes sense data and send to their respective CHs which aggregate and fuse the data, thus saving the energy as global communication is reduced due to local compression.

Once the CH receives data from its nodes it aggregates and fuses the data into a small set and sends to BS. Unbalanced energy consumption among the sensor nodes may cause network partition and node failures where transmission from some sensors to the sink node becomes blocked. Therefore, construction of a stable backbone is one of the challenges in sensor network applications.

LEACH [3] proposes a clustering based routing protocol for homogenous networks in which a node becomes CH by a probabilistic equation and forms a cluster of those nodes which receive strong signal to noise ratio from it. The nodes sense the environment and send data to CH where it is aggregated and finally send to BS. In LEACH there is a localized coordination amongst the nodes for cluster set up and locally compress the data to reduce global communication. CHs in LEACH are rotated randomly. Heterogeneous networks are more stable and beneficiary than homogenous networks. A number of protocols like SEP, DEEC and Threshold Distributed Energy-Efficient Clustering protocol (T-DEEC) have been proposed for WSNs. SEP [4] has two level of heterogeneity. In DEEC [5], CH selection is based on the ratio of residual energy and average energy of the network. The high energy nodes have more chances to become CH. In this way the energy is evenly distributed in the network. These routing protocols have some limitation due to their design and performance.

\section{The ACH Scheme}

\subsection{Optimal Number Of CHs}

The optimal probability of a node to take part in election for selection of CHs is a function of the spatial density when the nodes are uniformly distributed over the sensors' field. When the total energy consumption is minimum and energy consumption is well distributed over all sensors, the clustering is then called optimal clustering. The energy model we use for our simulation effect the optimal number of CHs. We use similar energy model as proposed in LEACH, SEP and DEEC. 
We have been giving particular attention to distribution of CHs in network so as energy in the network. Once nodes are deployed in region of interest the nodes locally coordinate for cluster set up and operation. Each node decides whether to become a  CH or not. The node generates a random number and compares it with the threshold value. If the number generated is less than or equal to the threshold value and the node has not been CH for the last $\frac{1}{p}$ round the node is marked as to be one of the CH. p is the probability of a node to become CH. In ACH scheme, once the CHs have been formed the CHs send a confirmation message to one another using CSMA-MAC protocol. The CH which receives a strong SNR from its adjacent CH will be marked as a normal node. For simplicity in our simulations we replace SNR by distance. We assume the CH will be made unmark and will become a normal node if its distance from the nearest CH is less than $12m$. The distance between CHs less than $12m$ is shown by "a" otherwise "b". After confirmation of CHs, nodes receive an association message from CHs and respond according to the strength of SNR. The clusters are thus formed and the CHs are well distributed in the network. This makes clusters even in terms of number of nodes in each cluster. In this way energy of CH dissipated in each round is comparably equal. The distant CHs' network is shown in fig. 2. We implement ACH scheme on LEACH, SEP, TEEN and DEEC. Simulation results show that ACH scheme performs better with all of 4 selected protocols. We initialize parameters for simulation, randomly deploy our nodes and start network's operation. In network's operation each node is checked whether eligible to become CH. We call this operation for epoch. If a node successfully pass through this test energy of the node is checked. Next comes the turn of CH formation. A node becomes CH if it has energy and is eligible to be CH which is shown in CH formation block. The CH then goes through neighbors' association phase and data transmission phase as shown in fig. 1.

\begin{figure}[h]
\centering
\includegraphics[height=7cm,width=9cm]{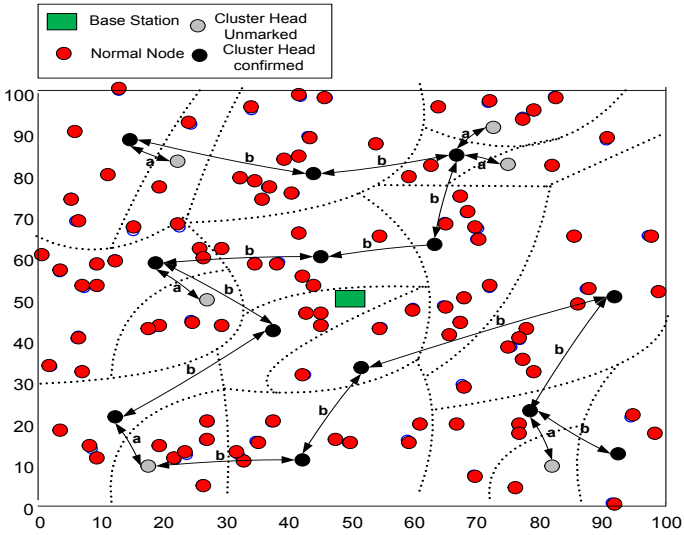}
\caption{ACH Scheme}
\end{figure}

\begin{figure}[h]
\centering
\includegraphics[height=7cm,width=8cm]{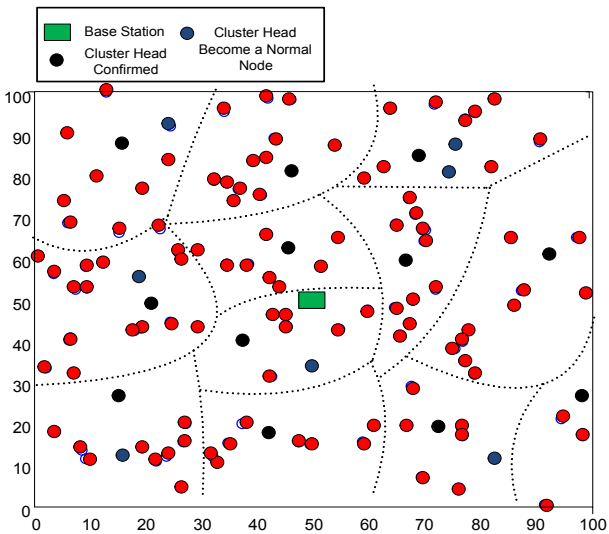}
\caption{Distant CHs}
\end{figure}

\section{Simulations}
We have implemented our protocol on Matlab [6] to evaluate its performance with LEACH, SEP, TEEN and DEEC. We have proposed LEACH-ACH, SEP-ACH, TEEN-ACH and DEEC-ACH. Our goals in conducting the simulation are as follows:

\begin{itemize}
\item Compare the performance of LEACH, SEP, TEEN, DEEC and their ACH versions on the basis of longevity of the network.
\item Compare throughput of LEACH, SEP, TEEN, DEEC and the respective ACH schemes.
\end{itemize}

We have performed our simulation on $100$ nodes and a fixed BS located in the center of the field. We randomly distributed $100$ nodes in a $100m×100m$ field. The most distant node from BS is at $70.7m$. The nodes have their horizontal and vertical coordinates located between $0$ and maximum value of the dimension which is $100$. All the nodes have different energies as the environment is heterogeneous. We simulate our protocol on the basis of initial energy as follows:

\begin{itemize}
\item The maximum energy of a node in the field is not more than 0.5J/bit.
\end{itemize}

The parameters used in our simulation are summarized in Table. 1.

\begin{table}
\centering
\caption{Parameters used in our simulations}
\begin{tabular}{|l|l|}
\hline
Parameter & Value  \\\hline
Simulation Area  & 100m 100m \\\hline
Location of BS & (50m, 50m) \\\hline
Number Of Nodes & 100 \\\hline
Initial Energy of Nodes (Maximum Value) & 0.5 J/node \\\hline
Packet Size & 4000 bits\\\hline

\end{tabular}
\end{table}

To analyze and compare the performance of our protocol with LEACH, SEP, TEEN and DEEC we have used two metrics. They are:

\begin{itemize}
\item Total number of dead nodes:

 This metric show the overall lifetime of the network. It gives us an idea about the stability period and instability period. This metric is an indication of the number of dead nodes with time.
 \item 	Through put:

 This metric is an indication of the rate of packets sent to BS.
\end{itemize}

\subsection{Implementing ACH Scheme on LEACH, TEEN, DEEC, and SEP}

\subsubsection{LEACH-ACH}
LEACH is a clustering based routing protocol for homogenous networks. In LEACH probability is same for all nodes to become CHs. The nodes compare the random value generated at each round with the threshold equation and become CH if the threshold value is less than the random number. The threshold formula is given by:

$
T(\emph{n})=\begin{array}{cc}
  \{ &
    \begin{array}{cc}
      \frac{P}{1-P(rmod\frac{1}{P})} & n\in G \\
    0 & {else}
    \end{array}
\end{array}
$

We implement ACH scheme in LEACH and found its results better than LEACH as shown in fig. 3.

\begin{figure}[h]
\centering
\includegraphics[height=6cm,width=8cm]{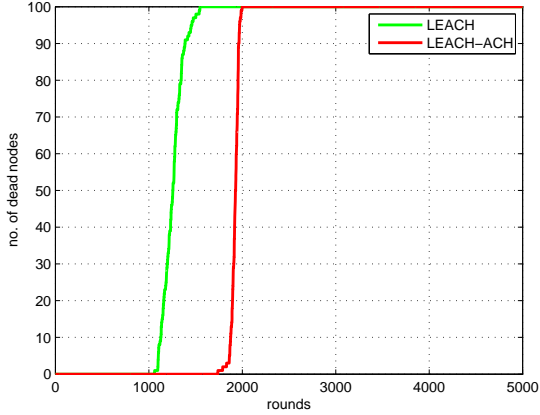}
\caption{LEACH Vs. LEACH-ACH}
\end{figure}

\subsubsection{TEEN-ACH}
TEEN is a routing protocol for reactive networks. In TEEN two thresholds: hard and soft have been introduced to reduce number of communications. After deployment of nodes CH set up phase starts in which CHs are formed. Once the CHs are confirmed the nodes sense environment and on their transmitter. When the the sense value reach hard threshold the transmitter is on and data is send to CH. This value is stored in an internal variable. Next time when the sense value reach the hard threshold, difference of stored value and sense value is obtained, if this value is greater or equal to soft threshold transmission is done otherwise transmitter are kept off. In TEEN-ACH we make CHs distant which makes energy dissipation even among CHs and thus very less energy is consumed in each round. Fig. 4 shows the behavior of TEEN against TEEN-ACH.

\begin{figure}[h]
\centering
\includegraphics[height=6cm,width=8cm]{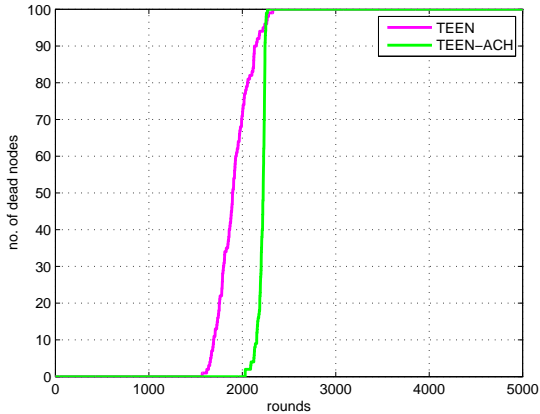}
\caption{TEEN Vs. TEEN-ACH}
\end{figure}

\subsubsection{DEEC-ACH}
DEEC-ACH is an extension of DEEC protocol which enhances stability period of DEEC. The CH selection criterion is based on DEEC protocol however we introduce ACH scheme which enhances the performance of DEEC.

DEEC is a heterogeneous routing protocol in which nodes have different initial energy as the network starts. DEEC uses initial and residual energy level of nodes to form CHs. Once sensor nodes are deployed in the region they locally coordinate for cluster set up and operation. Let $\emph{ni}$ denotes the number of rounds for which the node $\emph{Si}$ is CH often referred as the rotating epoch. $\emph{Popt}$ is our desired percentage of CHs and $\emph{ni}=1/\emph{popt}$ is the rotating epoch. By epoch we means that a node once becomes CH will not take part in CH formation for the next $1/\emph{popt}$ rounds. As in DEEC nodes have different energy levels the CH selection probability is different for each node and we call it average probability $\emph{pi}$. $\emph{pi}$ of nodes with more energy is greater. The average energy of network denoted by $\overline{E(r)}$ is given by eq.(1):

\begin{equation}
    \overline{E}(r)=\frac{1}{N}\sum_{i=1}^{N}E_i(r)
\end{equation}

The average probability of CHs per round per epoch is represented in eq.(2):

\begin{equation}
    \emph{pi}=\emph{popt}\,\,[1-\frac{\overline{E}(r)-Ei(r)}{\overline{E}(r)}]
\end{equation}

\begin{equation}
\sum_{i=1}^{N}pi=\sum_{i=1}^{N}popt\frac{Ei(r)}{{\overline{E}(r)}}=popt\sum_{i=1}^{N}\frac{Ei(r)}{{\overline{E}(r)}}=Npopt
\end{equation}

Eq.(3) shows the optimal number of CHs we want to achieve. The probability of nodes in the network to become CHs is based on the ratio of their residual energy and average residual energy of the network. The probability equation for nodes to become CH is given by eq.(4):

\begin{equation}
pi=\frac{potp N (1+a) E_i(r)}{(N+\sum_{i=1}^{N}a_i)\overline{E(r)}}
\end{equation}

Where $\emph{popt}$ is the desired percentage of CHs, $N$ is number of nodes, $\emph{Ei(r)}$ is residual energy of a node and $\overline{E}(r)$ is network's average energy. $Pi$ is average probability of a node to become CH. Each node creates a random number for itself and compares it with threshold equation, if the number generated is less than or equal to the threshold value the node is selected as CH for that round. The threshold equation is given by:

$
T(\emph{Si})=\begin{array}{cc}
  \{ &
    \begin{array}{cc}
      \frac{pi}{1-pi(rmod\frac{1}{pi})} & Si\in G \\
    0 & {else}
    \end{array}
\end{array}
$

Where $G$ represents the set of nodes eligible to take part in CH selection at round $r$. $\emph{Si}\,\,\,$$\epsilon$ $G$ consists of all those nodes which have not been CHs for the most recent $ni$ rounds. Once a node becomes CH it sends a confirmation message to all CHs. As soon as the CH is confirmed it sends an association message to all the nodes using CSMA-MAC protocol. The nodes respond according to strength of SNR received. The nodes associate themselves to that CH whose SNR is stronger. The CH then allocates TDMA slot to each node in the cluster. The nodes sense the environment and send data to their respective CHs in the TDMA slots allocated to them.

CH formation depends on random number generated, threshold value and energy of nodes. At some stages two or more very close (or intersecting) nodes become CHs and energy dissipation is even more and unbalance. We make CHs away in our protocol. Once a node becomes a CH it virtually take part in election of next CH. The CH decides area of next node taking part in election for CH. We assume that no node can become CH in (15m, 15m). A node which has become CH at (5m, 5m) will force the nodes to be normal nodes in the area ((5+10)m, (5+10)m) . In this way the CHs are made distant and we are able to achieve $Npopt$ CHs each round. Fig. 5 shows the comparison of DEEC with DEEC-ACH.

\begin{figure}[h]
\centering
\includegraphics[height=6cm,width=8cm]{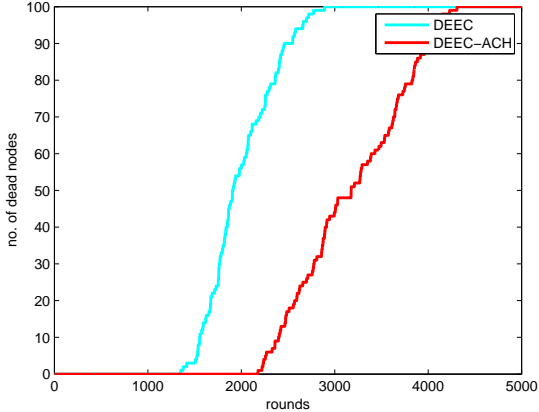}
\caption{DEEC Vs. DEEC-ACH}
\end{figure}

\subsubsection{SEP-ACH}

In this section we implement ACH scheme on SEP. In SEP we have two level heterogeneity. The normal nodes in SEP have $a$ times less energy than advance nodes. The probability of normal nodes in SEP differs from advance nodes as follows:

\begin{equation}
pnrm=\frac{p_{opt}}{1+a.m}
\end{equation}

\begin{equation}
padv=\frac{p_{opt}(1+a)}{1+a.m}
\end{equation}

where $pnrm$ in eq.(5) and $padv$ in eq.(6) is the probability equation for normal and advance nodes respectively. m is the fraction of advance nodes. Eq.(6) shows greater probability of advance nodes to become CHs. Each node in the network generates a random number for itself, compares itself with the threshold value and become CH if the number is less than or equal to threshold value. After the CH formation CH confirmation phase starts in which the CHs are made distant. We introduce ACH scheme in SEP which makes CHs formed in SEP distant. The energy in the nodes are thus conserved and the stability period of SEP is enhanced. Fig. 6 shows that SEP-ACH performs better than SEP. The $1st$ node in SEP dies at round $1130$ whereas in SEP-ACH the $1st$ node dies at round $2004$.

\begin{figure}[h]
\centering
\includegraphics[height=6cm,width=8cm]{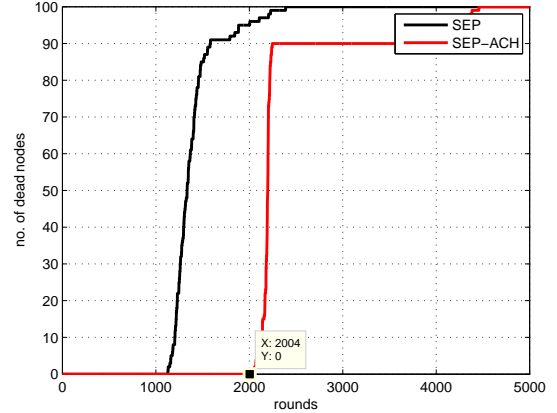}
\caption{SEP Vs. SEP-ACH}
\end{figure}

\section{Conclusion}
This paper deals with ACH scheme, a clustering technique for WSNs that enhances life time of LEACH, TEEN, DEEC, SEP and minimizing global energy consumption by distributing the load to all the nodes at different points in time. In DEEC-ACH high energy nodes are made CHs frequently than low energy nodes, making energy distribution evenly in the network. Also a CH take part in the selection of next CH thus the number of CHs are reduced and the CHs are made distant. We have forced those nodes which have not become CHs and are close to each other or intersecting. The energy of nodes are conserved in this way and stability period of the network is prolonged. DEEC-ACH outperforms LEACH as LEACH is not suited with heterogeneous environment. DEEC-ACH distribute the energy evenly in the network by giving high priority to high energy nodes in election for CHs and making CHs away from one another. DEEC-ACH also perform well than DEEC as the adjacent and very close nodes are made CHs in DEEC. This consume much energy of the nodes in the process of aggregation and fusion. The global communication is increased and the nodes die at a faster rate. The stability period and throughput of the network is decreased.

\end{document}